# 1

# ANALYSIS OF NANOSCIENCES AND NANOTECHNOLOGY AND THEIR APPLICATIONS

*José António Filipe[1] and Manuel Alberto M. Ferreira[2]*

[1]Department of Mathematics, ISTA—School of Technology and Architecture, University Institute of Lisbon (ISCTE-IUL), Information Sciences, Technologies and Architecture Research Center (ISTAR-IUL), Business Research Unit-IUL (BRU-IUL), Avenida das Forças Armadas, 1649-026 Lisboa, Portugal
E-mail: jose.filipe@iscte-iul.pt
ORCID: http://orcid.org/0000-0003-1264-4252

[2]Department of Mathematics, ISTA—School of Technology and Architecture, University Institute of Lisbon (ISCTE-IUL), Information Sciences, Technologies and Architecture Research Center (ISTAR-IUL), Business Research Unit-IUL (BRU-IUL), Avenida das Forças Armadas, 1649-026 Lisboa, Portugal
E-mail: manuel.ferreira@iscte-iul.pt
ORCID: http://orcid.org/0000-0002-4931-3281

**Abstract:**

The small world of matter is getting smaller and smaller. Nano sciences in recent years had huge developments allowing nanotechnologies to take enormous steps in the development of materials and processes. Numerous applications in a wide scope of fields are very beneficial for humans, and many researches in development are very promising. Applications in medicine, industry, electronics, energy, or aeronautics are only some examples of areas where enormous benefits exist, and potentialities are clear. Some nanotechnologies are already applied and others are in development or testing phases.

## 1.1. Introduction

We begin our chapter as in Ferreira and Filipe (2018) starting point: "beginning in mili ($10^{-3}$), then micro ($10^{-6}$) and now nano ($10^{-9}$) was reached in the miniaturization route. The "small" is becoming smaller and smaller. This is an evidence in a lot of domains and with great impact in our lives as it happens with this technology, which maybe paradoxically has the greatest one. In short, "small is beautiful". Briefly, of course, pico ($10^{-12}$), then femto ($10^{-15}$) will arrive. However, for now, stand in nano world".

Nanotechnology is a relatively recent topic in scientific research developments. However, central concepts exist over a longer period.

In 1959, Feynman was who first spoke on the opportunity of applying nanotechnology, in a talking at California Institute of Technology (Caltech).

In 1974, the term nanotechnology was created by Japanese Professor Norio Taniguchi, while working on the development of ultra-precision machines. He labelled precision machining with a tolerance of a micron or less.

The innovations of the 80s in nanotechnology's domain allowed the observation of materials at an incomparable atomic scale. Very powerful computers made possible large-scale material systems' simulations, the study of their structures and properties, and the production of new materials.

After the invention of the scanning tunneling microscope in 1981 and the discovery of fullerenes in 1985, one of the exceptional discoveries in technologic areas considering the nanoscale was precisely the discovery of carbon nanotubes in 1991. By the end of the 20th century, nanotechnology initiatives significantly increased the nanotechnology strength; and new enormous systematic advances in this area began to be possible. In the early 2000s the debates around the potential implications of nanotechnology and the feasibility of the possible applications considering the molecular nanotechnology made this area gain strategic interest, moving governments to promote funds for its development.

The developments in the 21st century gave this area the consolidation and strength to become considered one of the most promising areas in the contemporary societies.

Investments in USA are enough to be aware of the dimension and importance of this sector. Alone USA invested more than 18 billion dollars between 2001 and 2013 through the NNI - National Nanotechnology Initiative intending to make this sector a driver of economic growth and competitiveness[1]. This importance is highlighted by Roco (2011): "as of 2009, this new knowledge underpinned about a quarter of a trillion USD worldwide market, of which about $91 billion was in US products that incorporate nanoscale components".

Today, among the enormous investments in the very different nanotechnology's fields, the examples of the micro-manufacturing, organic chemistry and molecular biology are some where very massive investments in nanotechnology are being made.

As was happening in other parts of the world, this ascending importance of nanotechnology also made Europe to face the challenge and develop the sector as a main goal for European Union. At European Commission site, this goal is well defined, and the development of nanotechnology was considered a priority. So, the Horizon 2020 program was aimed "to bridge the gap between nanotechnology research and markets, and to realize the potential contribution to sustainable growth, competitiveness, environment, highly skilled jobs and increased quality of life. A few barriers need to be addressed, in order to leverage large scale market introduction of innovative, safe and sustainable nano-enabled products"[2].

The manipulation of matter using nanotechnology (on an atomic, molecular, and supramolecular scale) allows to mold materials' structures in nanoscale, achieving extraordinary intrinsic properties, making possible new and revolutionary applications. The nanoscience is developed in many different scientific areas, as it is the case of engineering, chemistry, physics, biology, or materials science, or still in pharmaceutical, electronic, energy, textile, coatings, and paintings for example.

---

[1] https://www.iberdrola.com/innovation/nanotechnology-applications.
[2] https://ec.europa.eu/programmes/horizon2020/en/h2020-section/nanotechnologies.

From the development of nanoscience as a whole, materials science works a benefitted field of action allowing, through the use of nanotechnology, the provision of stronger, more robust, lighter, more reactive, more sieve-like, more durable materials, better electrical conductors, for example, with endless new properties and with remarkable applications in many fields. For instance, the graphene is a modified carbon, is harder than steel, lighter than aluminum and is almost transparent; nanoparticles are used in areas such as electronics, energy, biomedicine, and defense.

In chemistry there are incredible developments in multiple areas, such as sensors for detecting very small amounts of chemical vapors. A large set of elements can be included in nanotechnology-based sensors: carbon nanotubes, zinc oxide nanowires, palladium nanoparticles. As nanotubes, nanowires, nanoparticles are very small sized, a small quantity of gas molecules are enough to change the electrical properties of the sensing elements, allowing to detect a very small concentration of chemical vapors[3].

Nanotechnology requires complexity and involves enormously small technology at this nanoscale world. By manipulating and controlling materials on a nanoscale, nanotechnology gained an important dimension in the context of the vast spread of applications, namely in industries and in the context of new economy. As materials change their manifestation's form depending on the scale they are, new directions for this research perspective involve aspects that are specific for dealing with this level of complexity.

All over the world, from USA to Europe or Asia innumerous projects were developed for industries, including a very diverse kind of applications, as it is the case for example of areas as automobile, military, optical, food or electronic, at the same time as there are enormous advances in other areas as in the medicine or in the air and water quality, among many others.

The nanoscale properties and the creation of useful devices and processes are today's important components of economic progress. The development of nanotechnology's area has very visible and significant new benefits in the societies' progress and lifestyle.

This chapter proposes an analysis for nanoscience and nanotechnology, their applications, and the resulting benefits for societies.

**1.2. Nanoscience and nanotechnology Applications**

In this nanoworld, nanoscience studies extremely small things. Nanoscience studies the matter, the particles, the structures on the nanometers' scale (one millionth of a millimeter, the scale of atoms and molecules). Specific microscopes provide a glimpse into this small reality and allows the means to fabricate and manipulate nanoscale elements. The manipulation of materials on this scale allows to develop structures with different properties than the ones on the macroscale (the quantum mechanical effects become important on the nanoscale). Several kind of materials specific characteristics in terms of electrical, mechanical, or optical properties are got, which are determined by the way molecules and atoms assemble on the nanoscale into larger structures.

When the size of a material is reduced to nanometer range chemical, physical and biological properties of the material changes which are entirely different from the properties of their individual atoms, molecules, or bulk materials (Tomar, 2012)

---

[3] UnderstandingNano.com (1). Nanotechnology Applications.
https://www.understandingnano.com/nanotech-applications.html.

Developments on nanoscience make innovations possible in a very huge set of fields. The discovery of the small world of matter allowed by technology intensifies the possibility of great "adventures" in the progress of humanity's quality of life, considering all those innumerous fields.

By the end of the 20th century, in the years from 1998 until 2000, the fragmented fields of nanoscale science and engineering were brought together under a unified science-based definition for nanotechnology. This vision would allow a definition that is reproduced as follows.

*"Nanotechnology is the ability to control and restructure the matter at the atomic and molecular levels in the range of approximately 1–100 nm, and exploiting the distinct properties and phenomena at that scale as compared to those associated with single atoms or molecules or bulk behavior. The aim is to create materials, devices, and systems with fundamentally new properties and functions by engineering their small structure. This is the ultimate frontier to economically change materials properties, and the most efficient length scale for manufacturing and molecular medicine. The same principles and tools are applicable to different areas of relevance and may help establish a unifying platform for science, engineering, and technology at the nanoscale. The transition from single atoms or molecules behavior to collective behavior of atomic and molecular assemblies is encountered in nature, and nanotechnology exploits this natural threshold"* (Roco, Williams and Alivisatos, 1999). After the publication of this definition in 1999, it would be adopted the following year as an official document by the US National Science and Technology Council (NSTC).

After the consultation of many worldwide experts, the above definition was agreed in that period (1998–1999) and got at some degree an international acceptance. Previously there was a conceptually different vision, more focused on either small feature under a given size, ultra-precision engineering, ultra-dispersions, or creating patterns of atoms and molecules on surfaces (Roco, 2011). The definition of nanotechnology presented above would give space for guidance on nanotechnology discovery and innovation in the future much interdisciplinary and multi-domain activities (Roco, 2011).

For the U.S. National Nanotechnology Initiative nanotechnology is the understanding and control of matter at dimensions between approximately 1 and 100 nanometers where unique phenomena enable novel applications. Encompassing nanoscale science, engineering, and technology, nanotechnology involves imaging, measuring, modeling, and manipulating matter at this length scale (National Nanotechnology Initiative, cited in Ridge, 2018).

GAEU Consulting defines nanotechnology as an area of science and engineering where phenomena that take place at the nanoscale ($10^{-9}$m) are utilized in the design, characterization, production, and application of materials, structures, devices, and systems. Nanomaterials (materials with at least one dimension or aspect below 100nm) can be naturally occurring, such as smoke, soot, dust, or sand. Others have been used long before it was known that they are nanomaterials, e.g. silica and carbon black[4].

Nanotechnology encompasses science, engineering, and technology at the nanoscale, which is about 1 to 100 nanometers. a nanometer is small as one-billionth of a meter. For reference, a sheet of paper is about 100,000 nanometers thick. Nanoscale matter can behave differently than the same bulk material. For example, a material's melting point, color, strength, chemical reactivity, and more may change at the nanoscale (National Science and Technology Council, Committee on Technology (COT) - Subcommittee on Nanoscale Science, Engineering, and Technology (NSET), 2018).

---

[4] https://gaeu.com/artiklar/this-is-nanotechnology-one-of-the-fastest-growing-markets-in-the-world/.

Nanotechnology is affecting all aspects of life through innovations that enable, for example: (a) strong, lightweight materials for aerospace applications; (b) clean, accessible drinking water around the world; (c) superfast computers with vast amounts of storage; (d) self-cleaning surfaces; (e) wearable sensors and health monitors; (f) safer food through packaging and monitoring; (g) regrowth of skin, bone, and nerve cells for better medical outcomes; (h) smart windows that lighten or darken to conserve energy; and (i) nanotechnology-enabled concrete that dries more quickly and has sensors to detect stress or corrosion in roads, bridges, and buildings (National Science and Technology Council, Committee on Technology (COT) - Subcommittee on Nanoscale Science, Engineering, and Technology (NSET), 2018).

The nanotechnology, as nanoscience applications, leads to the use of new nanomaterials and nanosized components allowing useful and revolutionary products and devices.

As nanotechnology is used for the manipulation of matter (on an atomic, molecular, and supramolecular scale), materials' structures in a nanoscale will reach extraordinary properties, as previously seen.

Reinforcing the importance of the fields where nanotechnology is being used, we can highlight some areas where we intend to present some developments as examples of the enormous advances in the nanotechnology area. Such areas to be considered below in terms of examples of applications are, for instance, health and medicine, energy, industry (food safety, for example), electronics and information technology, military and homeland security, transportation, or environmental science which are only some of the many of the enormous variety of fields that can benefit from these advances.

As nanotechnology gained a privileged space in the world of science and engineering, it has allowed a significant change in the way researchers think and is key in future technologies and solutions. As referred above, in the following subsections we present some applications in several different fields. It is not our aim to show all kinds of applications, even because they are so many that it would be impracticable, but only to present the nanotechnology's benefits and potentialities in several fields, showing a set of applications resulting from these developments.

Next, we can see a set of these applications and possibilities in several fields (GAEU Consulting[5]).

- ➢ In Medicine area:

  - Cancer detection and diagnosis using nanomaterials for imaging and biomarker detection;

  - Drug delivery for targeted drug administration, e.g. mesoporous $SiO_2$ and Au.

  - Prevention and control using research with nanotechnology.

  - Nanobiosensors for management and control of diseases.

- ➢ In Food and Cosmetics area:

  - Food packaging for better shelf life, nanosensors for temperature monitoring, nanoclay for protection.

---

[5] https://gaeu.com/artiklar/this-is-nanotechnology-one-of-the-fastest-growing-markets-in-the-world/.

- Bioavailability of nutrients improved using nanocapsules.
- Nanosize powders increase supplement uptake.
- Additives in creams and sunblock, $TiO_2$ and $ZnO$.

➢ In Human and Environmental Health:
- Water purification nanomaterials added waterwells.
- Nanocapsules for pesticide delivery.
- Nanotechnology applied to textiles for improved protection, e.g. Ag.
- Nanotubes used to increase the efficiency of solar panels.

➢ In Technology and Industry area:
- Quantum Dots used in LED panels for screens.
- Additives to paints, coatings, and primers.
- Improved physico-chemical properties of construction materials.
- Lighter, stronger automotive materials.

**1.2.1. Medicine and Health**

One of the important application's fields of nanotechnology is health and medicine. In this area, health improvements are got with the enhancement of new discoveries in this nanotechnology world. We can deal today in the health area with the so-called nanomedicine, nanotherapy, or pharmaceutical nanotechnology, for example. Coccia, Finardi and Margon (2012) refer that trends of nanotechnology grew in chemistry and medicine after the technological applications of new nanomaterials mainly in Chemical Engineering, Biochemistry, Genetics and Molecular Biology.

Nanotechnologies in the medicine area are being strongly developed for the diagnosis, prevention, and treatment of diseases. Medicine may benefit much from the development of nanoparticles or nanorobots to be used for example to make repairs on the cellular level. The delivery of small molecules, the therapeutic proteins or peptides, the creation of materials acting as nanocarriers to specific tissues or across biological barriers are examples of very challenging possibilities for the applications of nanotechnology to the health area.

Nanomedicine is the application of nanotechnology in medicine. Nanomedicine draws on the natural scale of biological phenomena to produce precise solutions for disease prevention, diagnosis, and treatment (see Nano.gov – National Nanotechnology Initiative). Earlier diagnosis, more individualized treatment options, and better therapeutic success rates may be possible with better imaging and diagnostic tools allowed by nanotechnology. Such kind of actions are largely beneficial in medicine diagnosis and treatment.

A promising area in nanomedicine is, for example, a method that allows that nanoparticles, made so small as molecules, may deliver drugs directly to cells in a human body. Gold nanoparticles have been adapted in commercial applications as probes for the detection of targeted sequences of nucleic acids.

It is possible to make more effective the patients' cancer treatment, in patients treated with chemotherapy, by reducing cells damaging[6]. Gold nanoparticles are being clinically investigated to serve as potential treatments for cancer and other diseases.

In the field of cardiovascular area, nanotechnology is also a very promising area. Follow some examples of investigation projects that have been carried on in this field.

In North Carolina State University, USA, some researchers are developing a method to deliver cardiac stem cells to damaged heart issue, being one of the examples where the research on this field is ongoing. The cardiac stem cells will attach nanovesicles, that are attracted to an injury to the stem cells for increasing the number of stem cells delivered to the injured tissue[7].

Also, researchers of the University of Santa Barbara, USA try to develop a nanoparticle to deliver drugs to plaque on an artery wall. They intend to attach a peptide protein to a nanoparticle, which will bind with the surface of the plaque. With this process, they try that these nanoparticles create an image of the existing plaque and deliver a drug to treat the illness (Castillo, August 7, 2018).

Another promising field on heart repair tissues is the research at the University of South Carolina and Clemson. Researchers are attempting to solve complications with collagen in the heart. If valves have too much collagen, they become stiff, and if they have too little, they get floppy. To work correctly, valves need adequate collagen. This nanotechnology research involves combining gold nanoparticles with collagen to modify the mechanical properties of the valves. This repair possibility of the defective heart valves allows to avoid the surgery (Castillo, August 7, 2018).

As seen above the diagnosis and treatment of diseases have significant benefits allowed by nanotechnology. The atherosclerosis, or he buildup of plaque in arteries are examples of that. In one technique, researchers created a nanoparticle that mimics the body's "good" cholesterol, known as HDL (high-density lipoprotein), which helps to shrink plaque (Nano.gov – National Nanotechnology Initiative).

Around the world, universities and institutes are studying several types of nanotechnological techniques and applications to a large set of diseases/health problems. Considering the huge number of applications, nanoengineering may make possible, for example, the development of gene sequencing technologies to enable single-molecule detection at low cost and high speed with minimal sample preparation and instrumentation; may also allow the use of regenerative technologies to be used in bones, neural tissues, or dental applications. The use of nanotechnology may contribute to the grow of tissue to allow human organs transplant, or to the use of graphene nanoribbons for repairing spinal cord injuries; preliminary research shows that neurons grow well on the conductive graphene surface; also the field of vaccines is being worked using nanotechnology (Nano.gov – National Nanotechnology Initiative).

**1.2.2. Energy**

In energy area, nanotechnology has already made remarkable progresses, what allows much effective performances and costs reduction.

---

[6] UnderstandingNano.com (2). Nanotechnology in Medicine - Nanoparticles in Medicine. https://www.understandingnano.com/medicine.html.

[7] UnderstandingNano.com (2). Nanotechnology in Medicine - Nanoparticles in Medicine. https://www.understandingnano.com/medicine.html.

Several examples of applications of nanotechnology in this area can be given (see for example, Ferreira and Filipe, 2018). We will present here some applications in different energy areas, as in fuel cells, solar cells, or batteries, for example.

In energy sector, i.e., nanomaterials technology intervenes at a number of stages of the energy flow that starts from the primary energy sources and finishes at the end user. There are just a few examples that prove that the limitation in non-renewable energy sources (oil, gas, coal and nuclear) can be solved by technological developments aimed at increasing efficiency and reducing emissions of renewable energy sources. These solutions in both energy and other sectors in general, require overcoming a few technology limitations, for which nanotechnology brings a unique opportunity (Serrano, 2010).

Fuel cells have benefited much from the developments of nanoscience and nanotechnology. The reduction of costs of catalysts that are used in fuel cells to produce hydrogen ions from fuel (as methanol, for instance) is being got. These catalysts (as it is the case of platinum, the most used on these processes) can be very expensive. For this process, what happens is that nanoparticles of platinum are being used, what allows the reduction of the cost of the process once the platinum quantity is reduced. Besides, nanoparticles of other materials (graphene coated with cobalt nanoparticles, for example) can be also used, what replaces totally the platinum with an evident reduction of costs. Also more efficiency is being got in membranes that are used in fuel cells in the separation process of hydrogen ions from other gases such as oxygen. In this situation, membranes contained in fuel cells allow hydrogen ions to pass through the cell, but not other atoms or ions, as the oxygen ones, which do not pass through. More efficient membranes are produced using nanotechnology making fuel cells more durable and agiler weight.

Also, in laptop computers or PDAs, nanotechnology can allow the replacement of batteries by small fuel cells (many using methanol), lasting longer than usual batteries. These fuel cells also allow to reinsert a new cartridge of methanol instead of the battery electrical recharge, saving for example the time for recharging it. Also, electric cars can benefit from the nanotechnological developments of fuel cells allowing to replace their batteries by these fuel cells. The most proposed fuel to these fuel cells, in fuel cells powered cars, is hydrogen.

Also, in the solar cells area, nanotechnology supports lower costs for new types of solar cells, either manufacture or installation costs.

Batteries are used in many areas, such as in transports, electronics, medical equipment, power tools, electricity storage. The increasing technological capacities to improve energy densities and to decrease charge times are passing now for the tremendous growing knowledge in nanotechnology area and its capacity for innovation. New nanotech batteries can allow these tremendous new advantages (such as the reduction of the time for recharge, the increase of the available power or reduction of risks). Nanotechnology can increase the safety in batteries, reducing or eliminating the chance of a short circuit.

**1.2.3. Industry**

Industry is an economic field where the nanotechnology is used, being the matter worked on a molecular or atomic scale. The developments in nanoscience and nanotechnology allow new and very innovative applications in the industry, in several different areas.

Indeed, nanotechnology proves to be very efficient by using very diverse techniques and methods on applications to a set of areas, as in the industry sector. Developments on nanotechnology allow to improve, and often completely transfigure, the industry production's methods and techniques, many of the used technology and many industry's sectors, allowing a very new and effective way of producing. Anyway, there is much research needed to certify these technologies. In the food industry, for example, there are yet further research required to study the effects resulting from the food processing technologies, food packaging materials, or food ingredients, that are promising fields on which nanotechnology is advancing and searching for potential benefits although it is being also studied potential adverse health effects.

In this area, Thiruvengadam, Rajakumar and Chung (2918) made an interesting study for the food industry. In particular, they analyzed advantages and risks of several materials and techniques for this industry. They show that "nanoemulsions display numerous advantages over conventional emulsions due to the small droplets size they contain: high optical clarity, excellent physical constancy against gravitational partition and droplet accumulation, and improved bioavailability of encapsulated materials, which make them suitable for food applications". Also showed that "nano-encapsulation is the most significant favorable technologies having the possibility to ensnare bioactive chemicals".

These authors reviewed the current nanotechnology research and its applications in food technology and agriculture, particularly nanoemulsion, nanocomposites, nanosensors, nano-encapsulation and food packaging. Given the state of the nanotechnology in this field, they proposed future developments in the developing field of agrifood nanotechnology and made an overview of nanostructured materials, and their current applications and future perspectives in food science.

### 1.2.4. Electronics and Information Technology

Electronics is one of the fields on which the effective advantages of nanotechnology are tremendously visible. The advances in computing and electronics allowed by nanotechnology are enormous, through for example smaller, faster, and more portable systems, a greater capacity of information's storage and management and with implications in a very large set of fields.

Transistors are now allowed to be much smaller. The speed of the changing process to "the smaller, faster and better" is very evident when considering the passage from the typical transistor's size of 130-250 nanometers by the beginning of this century to a one nanometer transistor in 2016. The magnetic random-access memory (MRAM) enabled by nanometer-scale magnetic tunnel junctions allow completely revolutionary properties in computers "boot" and data saving, for example. The more vibrant colors in televisions and ultra-high definition displays and their more energy efficiency are features got by nanotechnology, particularly by using quantum dots (see Nano.gov – National Nanotechnology Initiative).

In terms of assembly processes, innovations as the ones of "nanoparticle copper suspensions [which] have been developed as a safer, cheaper, and more reliable alternative to lead-based solder and other hazardous materials commonly used to fuse electronics in the assembly process" (see Nano.gov – National Nanotechnology Initiative).

Also "flexible, bendable, foldable, rollable, and stretchable electronics are reaching into various sectors and are being integrated into a variety of products, including wearables, medical applications, aerospace applications, and the Internet of Things". There are flexible electronics made possible using, for example, semiconductor nanomembranes for applications in smartphone and e-reader displays. Graphene and cellulosic nanomaterials are, by their turn, being used for various types of flexible electronics to enable wearable and "tattoo" sensors, photovoltaics can be sewn onto clothing, electronic paper can be rolled up. "Making flat, flexible, lightweight, non-brittle, highly efficient electronics opens the door to countless smart products".

Some other applications on computing and electronic products are including for example flash memory chips for smart phones and thumb drives; ultra-responsive hearing aids; antimicrobial/antibacterial coatings on keyboards and cell phone casings; conductive inks for printed electronics for RFID/smart cards/smart packaging; and flexible displays for e-book readers" (see again Nano.gov – National Nanotechnology Initiative).

Within the multiple and diverse applications in this area, in 2018, Osaka University-led researchers, in a joint research project with The University of Tokyo, Kyoto University, and Waseda University, constructed integrated gene logic-chips called gene nanochips. Using integrated factors on the nanochips, these self-contained nanochips can switch genes on and off within a single chip, preventing unintended crosstalk[8] (Allied Market Research).

**1.2.5. Military and Homeland security**

The security of nations, defense/military or homeland is a field on which nanotechnology will have a crucial role. In areas such as military, coast guard, police, fire, emergency response, or medical, the importance of nanotechnology is key in the future.

Nanoscience and nanotechnologies already made important advances in these areas. This relevance may be seen since the late 70s, when the US Department of Defense (DoD) initiated the Ultrasubmicron Electronics Research (USER) Program (the ultrasubmicron was related to project to size scales below micron; today it would be called only nanoelectronics research). Several programs followed this one, in the 80s, trying to develop and exploit the scanning tunneling microscopy and the atomic force microscopy, what allows today scientists to use "eyes and fingers" to the measurement and manipulation of nanometer sized objects. Also in the early 1990s DoD programs began to exploit nanostructures (Murday, 2003).

The importance of nanoscience and nanotechnology in terms of defense and homeland safety applications is so great that it can potentially change in some way the international balance of nations' power. Nanotechnology may be useful for example in the area of the development of broad range collectors and detectors of weapons of mass destruction, using for example nanomaterials including silica-based materials, molecular imprinted polymers, and silicon platforms (Reynolds and Hart, 2004); or in the preservation of properties in shipping storage by encoding information (Suciu, 2019).

In the past 20 years, metamaterials and artificial materials were engineered to bend electromagnetic, acoustic, and other types of waves in ways not possible in nature. A professor of electrical and computer engineering at the University of Arizona, Prof. Hao Xin, made a discovery with these synthetic materials allowing the possible building of microscopes with superlenses that see molecular-level details, or shields that conceal military airplanes and even people[9] (Homeland Security News Wire, January 27, 2015). These findings were published in Ye et al.. In the study, the authors experimentally demonstrate active metamaterial with simultaneous gain and negative index of refraction in microwave regime; and discuss the theory and implementation of an artificial microwave gain medium based on negative-resistance devices. Although these theoretical advances, metamaterials remain in study and in the testing phase; and potential great applications like invisibility cloaks actually will not appear on the market in the short time.

**1.2.6. Transportation**

Mathew et al. (2019), discuss the recent applications of nanotechnology in transportation field, which include nanofilters, GMR sensors, nanocatalysts, anti-glare coatings, carbon black in tyres, fuel additives, or dirt protection, for example. The authors present some interesting nanotechnological applications that are now in research and that are seen as viable in the future (nanosteel, low friction aggregate components, switchable materials, glare free wiper free glasses, environmental multisensors, situation adapted driving mode).

---

[8] https://www.alliedmarketresearch.com/nanotechnology-market.
[9] http://www.homelandsecuritynewswire.com/topics/nanotechnology.

Nanotechnology contributes crucially to necessary developments and the production of innovative materials and processes in the automotive, aerospace and water transportation sectors (Tomar, 2012).

Mansson et al. (2005) review the use of actin-based motors, (myosins; e.g., the molecular motor of muscle) in nanotechnology, discuss the importance of molecular motors devices responsible of cargo transportation in the cell and their employment in nanotechnological applications.

Tomar (2012) refers that modern tyres achieve their high mileage, durability and grip through nanoscale soot particles and silica. Materials with nanoparticles or layers at the nanoscales have beneficial effects on inner and outer surfaces, on the body or on the engine and drive. The author says that most of the researches and developments based on nanotechnology are in automobile sector. For details of important applications of nanotechnology in the transportation sector, see Tomar (2012): body parts, emissions, chassis and tyres, automobile interiors, electrics and electronics, engines, and drive trains.

Maurya (2018) shows how these developments are made in terms of the application of nanotechnology in automobiles. Nanotechnology is being applied to car body parts like emissions, chassis, tires, automobile interiors, electrics, etc. These body part applications include paint coatings, lightweight parts, self-cleaning and scratch resistant nanopolymers. The author also approaches the applications of nanotechnology in marine transportation; and concludes that by the application of nanotechnology, it is made possible to make transportations more efficient, smart looking, stronger, and durable.

Also, nanoscale sensors, communications devices, and other innovations enabled by nanoelectronics support an enhanced transportation infrastructure that can communicate with vehicle-based systems to help drivers maintain lane position, avoid collisions, adjust travel routes to avoid congestion, and improve drivers' interfaces to onboard electronics[10] (Allied Market Research).

### 1.2.7. Environment and Environmental Science

In 2003, a Workshop on Nanotechnology and Environment occurred in Arlington, V A, USA reporting the importance of this new field and its relevance for environment. This report approached the following topics: nanotechnology applications for measurement in the environment, nanotechnology applications for sustainable materials and resources, nanotechnology applications for sustainable manufacturing processes, nanotechnology implications in natural and global processes, nanotechnology implications in health and environment, infrastructure needs for R&D and education, ending with a series of recommendations for each area.

Today, environment is benefitting much from these nanotechnological applications and is being much developed through environmental science. The advances in science are particularly allowing effective applications of nanotechnology in environment protection.

Engineering science in environmental is going fast from one paradigm to another.

In Hussain and Mishra (Eds.) (2018), a large view on nanotechnology is given to show its importance in subjects related to the environmental science. Nanotechnology is expected to contribute for easier, faster and cheaper processes in environmental monitoring and remediation This book also presents an up-to-date information on the economics, toxicity and regulations related to nanotechnology and approaches the role of nanotechnology for a green and sustainable future.

---

[10] https://www.alliedmarketresearch.com/nanotechnology-market.

As previously seen, nanomaterials have tremendous and unique properties arising from their nanoscale dimensions. However, the effect of using these new materials on the health and environmental is still away from being completely known. Governmental agencies and research organizations are continuously working on the development of analyses and methods to evaluate the risks of manufacturing and using nanomaterials for both health and the general environment. In fact, we are far from an understanding of the environmental implications of the nanomaterials (Serrano, 2010).

Serrano (2010) presented an interesting overview of the progress on nanomaterials research and development that are closely related to their manufacture, properties, storage, transport, usage and release to the environment. The impact of each aspect of nanomaterials on the human body and the environment continues an open space to evaluate the balance between potential risks and rewards.

Nanotechnological products, processes and applications are expected to contribute significantly to environmental and climate protection by saving raw materials, energy and water as well as by reducing greenhouse gases and hazardous wastes. Using nanomaterials therefore promises certain environmental benefits and sustainability effects. However, nanotechnology currently may play a rather subordinate role in environmental protection, whether it be in research or in practical applications. Environmental engineering companies themselves attach only limited importance to nanotechnology in their respective fields, although potential environmental benefits are very reachable. Nanomaterials exhibit special physical and chemical properties that make them interesting for novel, environmentally friendly products[11] (Nanowerk).

### 1.2.8. Others/General

In Nano.gov – National Nanotechnology Initiative, it is possible to see a wide set of applications of nanotechnology in several areas for everyday uses ("everyday materials and processes"), their benefits, properties, the innumerous advantages for a large spectrum of nanoscale materials and processes, involving a large spectrum of kinds of uses and fields. Today there is an enormous number of researches and developments for applications in the nanotechnology area. We show in this study only some of these multiple possibilities of application. There are many organizations globally investing in nanotechnology market and its emerging applications.

To the ones in the above fields we add several in other areas. For example, nanoscale sensors and devices may provide cost-effective continuous monitoring of the structural integrity and performance of bridges, tunnels, rails, parking structures, and pavements over time.

All these factors are expected to be major nanotechnology market trends globally.

While nanotechnology is largely in development and in the research phase, there are engineered nanomaterials produced and already used currently in commercial applications. Some of the products that presently use nanotechnologies are for example sunscreens and cosmetics, longer-lasting tennis balls and light-weight, stronger tennis racquets, stain-free clothing and mattresses, polymer films used in displays for laptops, cell phones, digital cameras; coatings for easier cleaning glass, bumpers and catalytic converters on cars, protective and glare-reducing coatings for eyeglasses and cars[12].

### 1.3. Nanotechnology and Anti-Commons

---

[11] https://www.nanowerk.com/nanotechnology-and-the-environment.php.
[12] https://www.osha.gov/dsg/nanotechnology/nanotech_applications.html.

Heller (1998) has made the "tragedy of the anti-commons" a popular expression since the problem has been presented by Michelman (1982). Since then, the anti-commons theory has been used to a large set of situations and debated in several academic disciplines. It has been discussed within the property law area and assorted boundaries have been studied for the concept (Filipe, 2016).

As stated in Filipe, Ferreira and Coelho (2008), when many agents may take decisions over a resource, hold together and exploited by all parts, existing the possibility of imposing a veto decision, an anti-commons problem may rise. All agents have to agree about the resource's utilization. Otherwise, a situation of non-utilization for the resource or its underuse is possible. About anti-commons theory and examples see also, for instance, Coelho, Filipe and Ferreira (2010), Filipe, Ferreira and Coelho (2011), Filipe et al. (2012), Filipe (2015, 2016).

In Heller and Eisenberg (1998), it is suggested that excessively fragmented intellectual property claims could result in an anti-commons setting.

After the studies of Heller (1998) and Heller and Eisenberg (1998), it was established that broad patents may not lead to an efficient development of research and technology. In such a situation a problem of "anti-commons" may stand up – see also for example Eisenberg (2019), who refers that an anti-commons is a fragmented allocation of property rights in which resources are prone to underuse because it is costly to assemble necessary permissions to put resources to use. The more rights holders and the more varied their entitlements, the more challenging it is to avoid waste through bargaining. The patent system continuously creates new rights for new claimants, with limited opportunity to establish consensus valuations as technology changes.

An anti-commons involves situations for which the knowledge gathered by exclusive intellectual property rights might go unused because of the transactions costs of negotiating the necessary agreements among a multitude of owners, with divergent interests, and incompatible expectations about the values of their intellectual property (Bruns, 2000).

Nanotechnology as much as anti-commons are relatively new areas in science, only with developments in the last few decades. Accordingly, the issue of anti-commons involving nanotechnology has not been yet much worked in scientific terms although it is promising due to the complexity it involves and the high expected benefits it may allow.

As seen, nanotechnology requires very specific and complex information and involves very different areas such as chemistry, biology, physics, materials science, or engineering. Considering its originality and capacity for dealing with completely different materials with amazing capabilities and properties, applications must face an enormous field of different technologies and entities.

In the most recent decades, the development of nanotechnology, its applications, and the high profitability expectations, made that nanotechnology patents have been a priority for many who developed them. Reynolds (2009) states that the "nanotechnology gold rush has specifically targeted nanomaterials, nanotechnology's building blocks. Many of the patents that have been granted for nanomaterials are broad, general patents encompassing basic research. A driving force behind the patenting of basic research in nanotechnology was the development-oriented approach to patent rights". As seen before, the existence of broad-spectrum patents may inhibit innovation and efficiency in resources utilization.

Some proposals have been made for solving, at least partially, the problem of anti-commons when patents are involved. However, these possible solutions are not easily implementable. Some authors suggested possible solutions as Heller or Eisenberg. It is also the case of the following proposal in nanotechnology. After studying if the broad patenting of nanomaterials has led to the creation of an anti-commons in US, Reynolds (2009) proposed the adoption of a strict utility requirement as a solution to the problems posed by the tragedy of the anti-commons in nanotechnology in the US.

**1.4. Nanotechnology and Ethics**

In a previous work, we dealt extensively with ethics and social implications in terms of nanoethics. In this section we do not intend to develop this subject in an extensive way. By the contrary, we only intend to show that nanotechnologies involve risks that have to be taken account. In fact, nanotechnology involves many benefits, but also many potential risks. The experience teaches from situations like the ones arisen from, for example, the genetic modified organisms or the issues connected with the human genome that an ethical debate is necessary and must exist. Social and ethical questions in this area need to have an essential continuous discussion when nanotechnology is used. Ethics matters and requires consideration when the borders of nanotechnology applications are debated being central for policymakers but not exclusively, also for all the parties involved, including scientists, engineers, regulators.

This issue is very easily understood when a series of situations are presented.

For example, in privacy and national security issues, the invisibility cloaks for hiding objects or people trigger important evident questions on this area. Also bionic hornets arise serious problems and allow to concern about arms race or misuse by terrorists.

The use of nanotechnology for clothing, sports equipment, cosmetics, food, packaging, among others may bring important risks for people's health, or even for animals and environment.

The applications in medicine may also bring important discussion involving the borders of its use. For example, medical treatments may deal with moral and ethical serious issues that are not easy to deal with in certain circumstances. The extension of life span allowed by nanotechnology may bring to discussion a set of moral and ethical issues, involving for example the possibility of questioning about overpopulation's problems or pensions system sustainability.

In medicine area, there is an interesting question to answer. Considering that the use of nanotechnology in medicine may bring the possibility to wealthy individuals who can afford to buy changes and modify themselves physically while others will not be able to afford to do it, what makes poorer people remaining the same state. Besides, there is the question to put about the right people have to change themselves.

We can imagine the innumerous moral and ethical situations emerged from nanotechnology applications on human-enhancing capacities involving, for example, cybernetic body parts or an exoskeleton to give a superhuman strength or infrared vision. Also, one day, nanotechnology may have a role in extending the human lifespan to the point of near-immortality with all the political, economic and social implications it has and the resulting ethical concerns.

Nanotechnology products, or nanomaterials (e.g., nanotubes, nanoparticles, nanofibers, nanowires, nanocomposites, and nanofilms) can be unsafe to human health because of the way they are manipulated and treated at near atomic scale, introducing concerns related to ethical, social, philosophical, environmental, biological, and other legal issues (Asmatulu, Zhang, Asmatulu, 2013). These materials may generate a set of uncertainties because of the lack of specific rules and regulations governing their manufacture and manipulation.

Once these technologies may be highly effective in several ways, questions on the impact it has on the jobs may be raised. As nanotechnology may allow to replicate the generality of things, the consequences are obvious in terms of the doubts it poses and the risks arisen from that.

### 1.5. Discussion and Final Considerations

Nanotechnology carries a significant impact, and serves as a revolutionary and beneficial technology across many fields, including industry, communication, medicine, transportation, agriculture, energy, materials and manufacturing, consumer products, households, etc. A large domain of uses and applications is expected to be crucial towards the growth of nanotechnology market size.

As nanotechnology deals with effective properties of materials on the nanoscale, they bring enormous potentialities to a large spectrum of applications.

In this chapter, we bring a set of applications, the benefits and the awareness of the risks that are yet unknown to discussion.

Anyway, the benefits and the uses that nanotechnology brought to the humanity are so large that nowadays nanotechnology is already one of the most challenging fields in science, existing in the more diverse areas of knowledge with applications in so different areas as medicine, energy, industry, or electronics, among many others.

The fragmentation of patents in nanotechnology area may induce some difficulties to the innovation in this field, in a problem defined as "anti-commons".

The development of the research in anti-commons associated to nanotechnology is seen as very interesting and deserves additional research. In our chapter the aim was only approaching it in the context of the illustrations we studied. For that reason, this issue is left for future work.